\documentclass{article}

\usepackage{arxiv}

\usepackage{url}            % simple URL typesetting
\usepackage[ruled,noline]{algorithm2e}
\usepackage{amsmath}
\usepackage{amsfonts}
\usepackage[colorinlistoftodos]{todonotes}
\usepackage{float}
\usepackage{url}            % simple URL typesetting

\usepackage{parskip}  % remove indents in paragraphs

\DeclareFontFamily{U}{wncy}{}
\DeclareFontShape{U}{wncy}{m}{n}{<->wncyr10}{}
\DeclareSymbolFont{mcy}{U}{wncy}{m}{n}
\DeclareMathSymbol{\comb}{\mathord}{mcy}{"58} 
\usepackage{accents}
\newlength{\dhatheight}

\title{Fast Variable Density Poisson-Disc Sample Generation with Directional Variation}

\author{
  Nicholas Dwork\thanks{www.nicholasdwork.com, nicholas.dwork@ucsf.edu} \\
  Department of Radiology and Biomedical Imaging \\
  University of California in San Francisco
    \And
  Corey A. Baron \\
  Robarts Research Institute \\
  The University of Western Ontario
    \And
  Ethan M. I. Johnson \\
  Department of Biomedical Engineering \\
  Northwestern University
    \And
  Daniel O'Connor \\
  Department of Mathematics and Statistics \\
  University of San Francisco
    \And
  John M. Pauly \\
  Department of Electrical Engineering \\
  Stanford University
    \And
  Peder E. Z. Larson \\
  Department of Radiology and Biomedical Imaging \\
  University of California in San Francisco
}

\begin{document}
\maketitle

\begin{abstract}
    We present a fast method for generating random samples according to a variable density Poisson-disc distribution.
    A minimum threshold distance is used to create a background grid array for keeping track of those points that might affect any new candidate point; this reduces the number of conflicts that must be checked before acceptance of a new point, thus reducing the number of computations required.  We demonstrate the algorithm's ability to generate variable density Poisson-disc sampling patterns according to a parameterized function, including patterns where the variations in density are a function of direction.  We further show that these sampling patterns are appropriate for compressed sensing applications.  Finally, we present a method to generate patterns with a specific acceleration rate.
\end{abstract}

% keywords can be removed
\keywords{Poisson Disc \and Parallel Imaging \and Compressed Sensing}

\section{Introduction}
\label{sec:intro}

In MRI, multiple coils and compressed sensing have both reduced the number of samples required to generate diagnostic quality imagery.  The multiple coils provide additional spatial encoding that is used to interpolate missing k-space data points (a technique commonly called \textit{parallel imaging}) \cite{sodickson1997simultaneous,pruessmann1999sense,griswold2002generalized}.  Compressed sensing takes advantage of the \textit{a priori} knowledge that most of the values of the image are approximately $0$ after a sparsifying linear transformation (e.g. a Daubechies Wavelet transform).  When the system matrix satisfies specific properties (e.g. the Restricted Isometry Principal, the Restricted Isometry Principal in Levels, or the Mutual Coherence Conditions) then the error on the final image is bounded \cite{donoho2003optimally,candes2006stable,adcock2017breaking}.  Remarkably, these conditions can often be achieved with a random sampling pattern \cite{candes2008introduction}.  Compressed sensing has been used in MRI with great success \cite{lustig2007sparse,vasanawala2011practical,gamper2008compressed}.

When combining multi-coil imaging with compressed sensing, one wants to employ a random sampling pattern to satisfy the compressed sensing requirements, but still keep samples far enough from each other to take advantage of the spatial encoding of the multiple coils.  A poisson-disc sampling pattern can be used to simultaneously satisfy these properties \cite{vasanawala2011practical}.  It is desirable that the pattern generation algorithm be fast in order to permit investigation of different sampling distributions and determine the advantages and disadvantages of each one.  Furthermore, it should accommodate densities that depend on direction to account for different coil configurations.  And finally, the sampling pattern should satisfy a desired overall acceleration factor.

The simplest dart throwing algorithm for generating a poisson disc sampling pattern (randomly choose a point, verify that the point is not too close to any existing point, repeat) is a slow process \cite{cline2009dart}.  Other methods are more efficient, but impose additional requirements (e.g. a mesh defining a surface of interest, or a tiling of the space where the density along the edges of the tiles may be noticeably different) \cite{lagae2008comparison}.  These are confounding effects that are not required for sample generation in MRI, where the region of k-space of interest is a simple rectangular subset of a Euclidean space.  In \cite{bridson2007fast}, Bridson described a fast $\mathcal{O}\left(n\right)$ algorithm for generating a poisson disc sampling pattern with a constant density.  In \cite{tulleken2008poisson}, Tulleken adapted this algorithm to accommodate a variable density sampling pattern based on \textit{a priori} knowledge of the largest poisson-disc radius parameter.  Tulleken's method is $\mathcal{O}\left(n^2\right)$.  Notably, this method cannot accommodate a sampling density that depends on direction.

In many cases of sample generation with MRI, we know the minimum distance between samples \textit{a priori}.  In this paper, we alter the methods of \cite{bridson2007fast,tulleken2008poisson} to take advantage of this knowledge.  We present a faster method of generating samples according to a variable density Poisson-disc distribution in a rectangular subset of a Euclidean space with an arbitrary number of dimensions.
We make three novel contributions for generating variable density poisson disc sampling patterns to be used in MRI.
\begin{itemize}
  \item We present a more computationally efficient and faster method than the state of the art.
  \item We present a computationally efficient method to accommodate rotationally asymmetric sampling density, permitting more acceleration in one direction than another.
  \item We present an automatic method to generate a sampling pattern with a desired overall acceleration rate, even with an asymmetric sampling density.
\end{itemize}

\section{Methods}
\label{sec:methods}

\subsection{Background}
\label{sec:background}

\begin{figure}[ht]
  \begin{center}
    \includegraphics[width=0.8\linewidth]{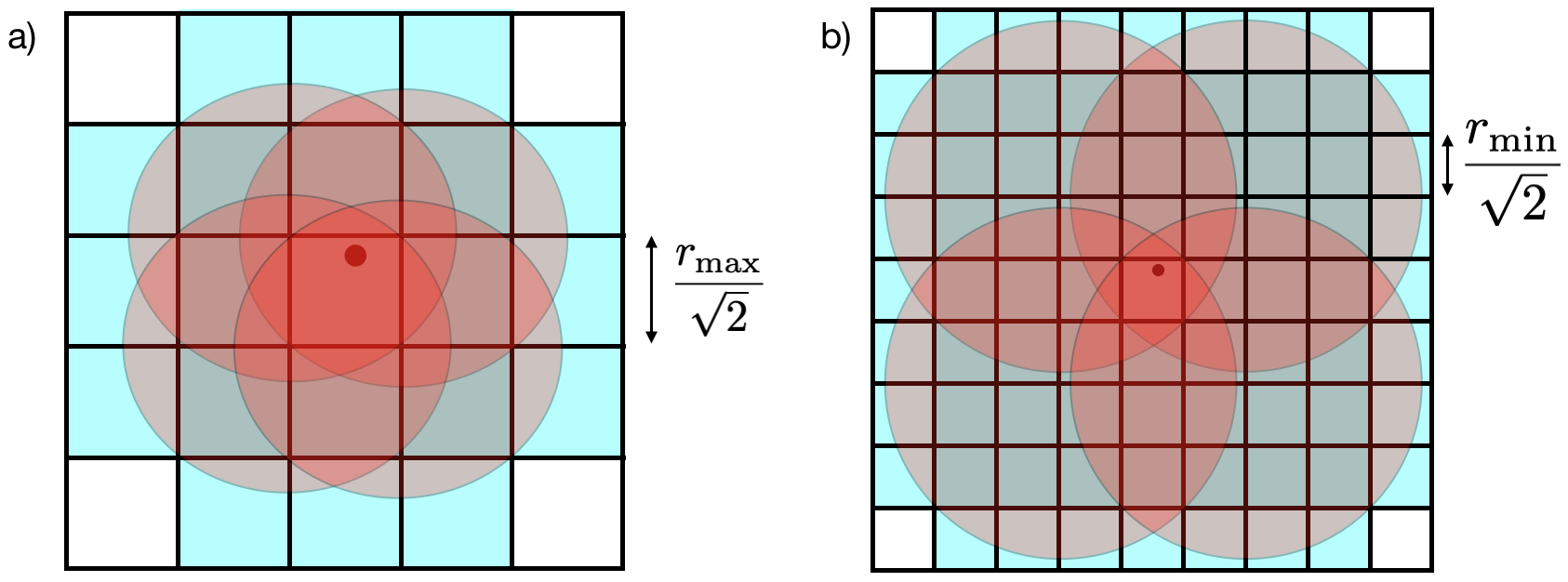}
  \end{center}
  \caption{ \protect\label{fig:nearbyPoints} a) For a new point (shown with a black dot), only those cells intersected by the red circles (shaded blue) can contain points close enough to violate the distance threshold.  If the point were in $n$ dimensions instead of $2$ dimensions, the length of each side of all cubes would be ${r_{\max}}/\sqrt{n}$.  b) A new candidate point $y_j$ with Poisson-disc parameter $r_j$ is illustrated with the black dot.  In this figure, $\text{ceiling}(r_j/{r_{\min}})=4$.  The red circles have diameter $4\,r_{\min}$ and are tangent to the corner of the cell containing $y_j$.  The blue squares are those cells that could contain points that are within $r_j$ of  $y_j$.  If the point were in $n$ dimensions instead of $2$ dimensions, the length of each side of all cubes would be $r_{\min}/\sqrt{n}$.}
\end{figure}

The method of \cite{bridson2007fast} reduces the computational time dramatically over the dart throwing algorithm by utilizing a background grid.  Assuming a fixed Poisson-disc parameter $r$, the method partitions the space into a set of cubes where the edges have length $r/\sqrt{d}$ (where $d$ is the number of dimensions of the space).  The cube edge is set to this length so that the cube's diagonal has length $r$; thus, each cell can contain at most $1$ point.  For each new point, then, one need not check the distance to all other points.  Instead, only those points that are indexed in grid cells within range (as illustrated in the Fig. \ref{fig:nearbyPoints}a, where $r=r_{\max}$) need to be checked \cite{tulleken2008poisson}.

The set of valid points is initialized by choosing a point $x_1$ at random in the space.  A list of point indices (called the active list) is initialized with index $1$.  An array of size equal to the background grid is created and initialized so that all values are $0$ (meaning that the corresponding cell does not contain any points).  The element of the background array that corresponds to the cell containing $x_1$ is set to $1$.  Computation then proceeds as detailed in Alg. \ref{alg:fastPD}.

\begin{algorithm}[ht]
    \protect\caption{Fast Poisson-Disc Sampling with constant $r$}
    \label{alg:fastPD}

    \textbf{Inputs:} $r$, active list, background array, parameter $k$

    \textbf{While: } active list is not empty

    \hspace{1em} Choose a point randomly from the active list, $x_i$

    \hspace{1em} Create $k$ new points uniformly at random in the spherical annulus 
    
    \hspace{2em} between radii $r$ and $2r$ centered on $x_i$

    \hspace{1em} \textbf{For each:} created point $y_j$

    \hspace{2em} Find the indices of those points that may be closer than $r$ to $y_j$
    
    \hspace{3em} using the background grid as illustrated in Fig. \ref{fig:nearbyPoints}a.  Find the
    
    \hspace{3em} distance between $y_j$ and the points with those indices.

    \hspace{2em} \textbf{If} the distance between $y_j$ and all existing points is greater
    
    \hspace{3em} than $r$
    
    \hspace{3em} Add $y_j$ to the list of points, add $j$ to the active list, and add
    
    \hspace{4em} index $j$ to the appropriate elements of the background grid.
    
    \hspace{2em} \textbf{End If}

    \hspace{1em} \textbf{End For}

    \hspace{1em} If none of the $k$ points were valid, remove $i$ from the active list.

    \textbf{End While}

    \textbf{Outputs: } List of points
\end{algorithm}

In Alg. \ref{alg:fastPD}, one must choose points randomly on a spherical annulus.  The number of points chosen is denoted by $k$.  In \cite{bridson2007fast}, Bridson suggests a value of $k=30$; we have found that the processing is faster with comparable results in two dimensions when $k=10$.  To choose points in two dimensions, construct a vector with angle (in radians) chosen uniformly at random on $[-\pi,\pi)$ and magnitude chosen uniformly at random on $[r,2r]$.  To sample a point at random in $n$ dimensions, generate a vector with $n$ elements where the value of each element is a realization of a normally distributed random variable.  Since the normal distribution is rotationally symmetric, every direction has equal probability density.  Then, scale this vector to a magnitude chosen uniformly at random on $[r,2r]$.

The method by Tulleken of \cite{tulleken2008poisson} accommodates a variable density Poisson-disc sampling pattern (meaning that the parameter $r$ changes as a function of location) by altering the background array so that 1) each element of the grid accepts a list of points and 2) the cell size must be computed from the maximum possible radius.  The index of each new point is added to the list of the background array element that corresponds to the grid cell that contains the new point.

Though this would result in a realization of the desired sampling pattern, it is inefficient, as the following thought experiment illustrates.  Suppose that $r$ is small near the center of the image and increases as the distance from the center increases.  In this case, the largest values of $r$ would be attained at the corners of the region.  Indeed, this is the most common use case for compressed sensing MRI applications.  These values could be so large that the rectangular region of interest would be divided into a small number of large grid cells, meaning that many points would be listed within each grid cell and all of those points would need to be checked with each new additional point.  The computational cost degenerates to that of the extremely slow dart throwing algorithm.  In section \ref{sec:fastAlgorithm}, we explain how to overcome this inefficiency.

\subsection{Fast algorithm}
\label{sec:fastAlgorithm}

For the fast algorithm, it is assumed that a positive minimum bound on $r$ exists: $r_\text{min} > 0$.  This is almost certainly the case with MRI.  The k-space samples need not be closer than the inverse of the field-of-view.  Moreover, with compressed sensing, the center region of k-space is often fully sampled (meaning that samples are separated by a distance equal to the inverse of the field-of-view) \cite{levine20173d,vasanawala2011practical}.  With these applications, the density of samples do not vary unless they are located some positive distance from the origin.  Thus, for any variable density scheme that reduces the sampling density as distance from the origin increases, the size of the fully-sampled center region can be used to determine $r_{\text{min}}$.

The region of interest is partitioned into a grid of cubes where the edges all have length $r_{\text{min}}/\sqrt{n}$.  Grid elements do not contain the indices of points that fall within their boundary.  Instead, each grid element contains a list of indices of those points that might have their threshold distance violated by a point in the grid cell.  When a new point is considered as a candidate for the sample distribution, its distance is checked against all of those points with indices located in its grid cell.  If the candidate point is not too close to any existing points, then its index is added to all of those grid cells where its distance threshold reaches (as illustrated in Fig. \ref{fig:nearbyPoints}b).

The set of valid points is initialized by choosing a point $x_1$ at random in the space.  Add index $1$ to the active list and add index $1$ to all those background grid cells that might contain points within the threshold distance of point $x_1$ according Fig. \ref{fig:nearbyPoints}b.  Then proceed according to Alg. \ref{alg:fastVDPD}.

\begin{algorithm}[ht]
    \protect\caption{Fast Variable Density Poisson-Disc Sampling}
    \label{alg:fastVDPD}

    \textbf{Inputs:} $r_{\min}$, active list, background array, parameter $k$

    \textbf{While: } active list is not empty

    \hspace{1em} Choose a point randomly from the sample list, $x_i$

    \hspace{1em} Identify the poisson-disc parameter for this point, $r(x_i)$

    \hspace{1em} Create $k$ new points uniformly at random in the spherical annulus 
    
    \hspace{2em} between radii $r(x_i)$ and $2r(x_i)$ centered on $x_i$

    \hspace{1em} \textbf{For each:} created point $y_j$

    \hspace{2em} Compute the distance between $y_j$ and those points with indices
    
    \hspace{3em} in the list contained in the background grid element that
    
    \hspace{3em} corresponds to the cell containing $y_j$.

    \hspace{2em} \textbf{If} the distance between $y_j$ and all existing points is greater
    
    \hspace{3em} than $r(x_j)$
    
    \hspace{3em} Add $y_j$ to the list of points, add $j$ to the active list, and add 
    
    \hspace{4em} index $j$ to the appropriate element of the background array.
    
    \hspace{2em} \textbf{End If}

    \hspace{1em} \textbf{End For}

    \hspace{1em} If none of the $k$ points were valid, remove $i$ from the active list.

    \textbf{End While}

    \textbf{Outputs: } List of points
\end{algorithm}

For the results presented in this work, we used the following function for the poisson-disc parameter $r$:
\begin{equation}
  r_{\gamma}(x) = \frac{ \|x\|_2 + 0.15 }{ \gamma },
  \label{eq:pdParameter}
\end{equation}
where $\|\cdot\|_2$ represents the $L_2$ norm.  For this function, $r_{\min}=0.15/\gamma$ and $r_{\max}=( \|x_{\text{corner}}\|_2 + 0.15 ) / \gamma$ where $x_{\text{corner}}$ is any corner of the sampling domain.

\subsection{Sampling density that is based on direction}

The rectangular region of k-space of interest is usually the $[-0.5,0.5]^n$ cube\footnote{Here, set multiplication is the Cartesian cross product.}.  (An exception would be a Homodyne sampling pattern, where the rectangular region is a little more than half of this cube \cite{noll1991homodyne}.)  As discussed, it may be desirable to alter the density distribution of points as a function of direction.  That is, in addition to the poisson disc parameter $r$ being a function of location, it would also be a function of radial direction to take advantage of coil placement geometry.  For a general function of radial direction, this is a computationally challenging task since it requires determining the direction between points and evaluating this function.  However, in MRI, it is usually the case that we are interested in acceleration rates that differ along the axes of the rectangular region.  For example, with a birdcage coil, the spatial encoding of each coil differs significantly in the transverse plane but differs little in the longitudinal direction.  Therefore, a higher acceleration is possible in the transverse plane than in the longitudinal direction.  And example of this can be seen in Fig. \ref{fig:kneeRecons}a.

Since we are interested in acceleration rates that differ along the axes of the rectangular region, a simple trick avoids determining the angle between points and evaluating the parameter function.  For simplicity, we will consider a two dimensional case.  Suppose one wants to undersample the second dimension by an additional factor of $\nu$ (meaning the density of samples will be greater in the first dimension than in the second).  Then, one can generate a variable density poisson-disc sampling pattern on the region $[-0.5,0.5] \times [-0.5/\nu,0.5/\nu]$ quickly with a radially symmetric Poisson-disc parameter using Alg. \ref{alg:fastVDPD}.  After the points are generated, scale the resulting pattern by $\nu$ in the second dimension to generate a pattern of points on $[-0.5,0.5] \times [-0.5,0.5]$.

\subsection{Specifying the Acceleration Rate}
The acceleration rate of a sampling pattern is the number of samples acquired divided by the number of samples required for full sampling.
It is often convenient to be able to specify an overall acceleration rate and attain a corresponding sampling pattern.  In this section, we describe a method to attain this goal.

In order to do so, we require a poisson-disc parameter function $r_{\gamma}:\mathbb{R}^d\rightarrow\mathbb{R}$ parameterized by $\gamma\geq 0$ such that $r$ decreases monotonically with increasing $\gamma$.  (Equivalently, the overall acceleration rate increases with increasing $\gamma$.)  Note that \eqref{eq:pdParameter} satisfies this property.
Then, with the computational efficiency of Alg. \ref{alg:fastVDPD}, we can now specify an overall acceleration rate $\alpha$ and determine the desired sampling pattern with a binary search algorithm in a reasonable amount of time.  (The total computational time is the time of any single iteration multiplied by the number of iterations in Alg. \ref{alg:binarySearch}.)

In order to use the binary search, one must supply bounds $\gamma_{\text{min}}$ and $\gamma_{\text{max}}$.  Since $\gamma$ is positive, $\gamma_{\text{min}}=0$ is a lower bound.  The $\gamma$ that corresponds to $r_{\text{min}}$ would be $\gamma_{\text{max}}$.  The complete search is specified in Algorithm \ref{alg:binarySearch}.

\begin{algorithm}[ht]
    \protect\caption{Binary Search to Find Pattern for Specified Acceleration Rate}
    \label{alg:binarySearch}

    \textbf{Inputs:} Desired acceleration rate $\alpha$, bounds $\gamma_{\min}$ and $\gamma_{\max}$, tolerance $tol>0$.

    \textbf{Do: } 

    \hspace{1em} $\epsilon = ( \gamma_{\max} - \gamma_{\min} ) / 2$

    \hspace{1em} $\gamma_{\text{mid}} = \epsilon + \gamma_{\min}$

    \hspace{1em} Determine the sampling pattern $P$ using Algorithm \ref{alg:fastVDPD} with
    
    \hspace{2em} parameter function $r_{\gamma_{\text{mid}}}=\alpha_{\text{mid}}$

    \hspace{2em} \textbf{If:} Acceleration rate $> \alpha$
    
    \hspace{3em} $\gamma_{\max} := \gamma_{\text{mid}}$
    
    \hspace{2em} \textbf{Else:} 
    
    \hspace{3em} $\gamma_{\min} := \gamma_{\text{mid}}$

    \textbf{While} $\epsilon > tol$

    \textbf{Outputs: } Pattern $P$
\end{algorithm}

\section{Results}
\label{sec:results}

\begin{figure}[H]
  \centering
  \includegraphics[width=0.95\linewidth]{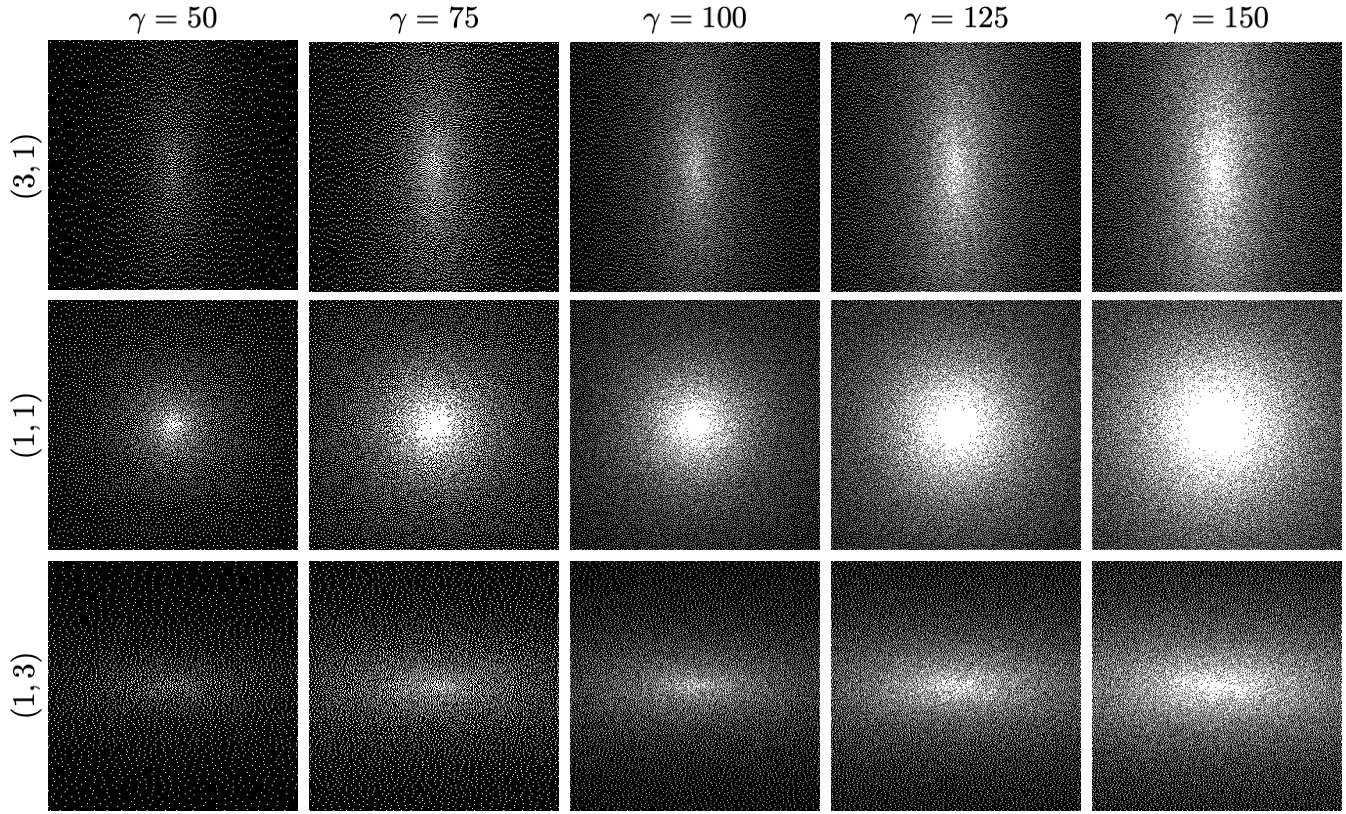}
  \caption{Variable density poisson disc sampling patterns generated with the fast algorithm of Alg. \protect\ref{alg:fastVDPD}.}
  \label{fig:vdpdSamplingPatterns}
\end{figure}

Figure \ref{fig:vdpdSamplingPatterns} shows variable density sampling patterns generated with Algorithm \ref{alg:fastVDPD} using the poisson-disc parameter function of \eqref{eq:pdParameter} for \\
$\gamma\in\{50,75,100,125,150\}$ and additional directional undersampling of 3.  As expected, as $\gamma$ increases, the overall acceleration rate (equal to the number of samples divided by the size of the domain) decreases.

Table \ref{tbl:runTimes} compares the time required to compute the variable density poisson disc sampling pattern with Alg. \ref{alg:fastVDPD} to the time required by the Tulleken algorithm for the sampling patterns shown in Fig. \ref{fig:vdpdSamplingPatterns}.  The improvement in runtime is between $30-50\%$.  The algorithms were implemented in C on a 2012 Macbook Pro with a 2.5 GHz Intel i7 processor.

\begin{table}[ht]
  \centering{}
  \begin{tabular}{|c|c|c|c|c|c|}
    \hline
     & \multicolumn{5}{|c|}{ Processing Time (ms):  Alg. \ref{alg:fastVDPD} / Tulleken } \\
    \hline
     & $\gamma=50$ & $\gamma=75$ & $\gamma=100$ & $\gamma=125$ & $\gamma=150$ \\ \hline
    (3,1) & 10 / 20 & 20 / 40 & 40 / 70   & 60 / 110  & 110 / 150 \\ \hline
    (1,1) & 20 / 30 & 50 / 80 & 100 / 140 & 150 / 220 & 230 / 340 \\ \hline
    (1,3) & 10 / 20 & 20 / 40 & 40 / 70   & 70 / 110  & 110 / 150 \\ \hline
  \end{tabular}
  
  \caption{ \protect\label{tbl:runTimes} Runtimes for generating variable density Poisson-disc sampling patterns with Alg. \ref{alg:fastVDPD} and the Tulleken algorithm.  Time is reported in milliseconds. In all cases, Alg. \ref{alg:fastVDPD} is faster by $30-50\%$. }
\end{table}

\begin{figure}[ht]
  \begin{center}
    \includegraphics[width=0.5\linewidth]{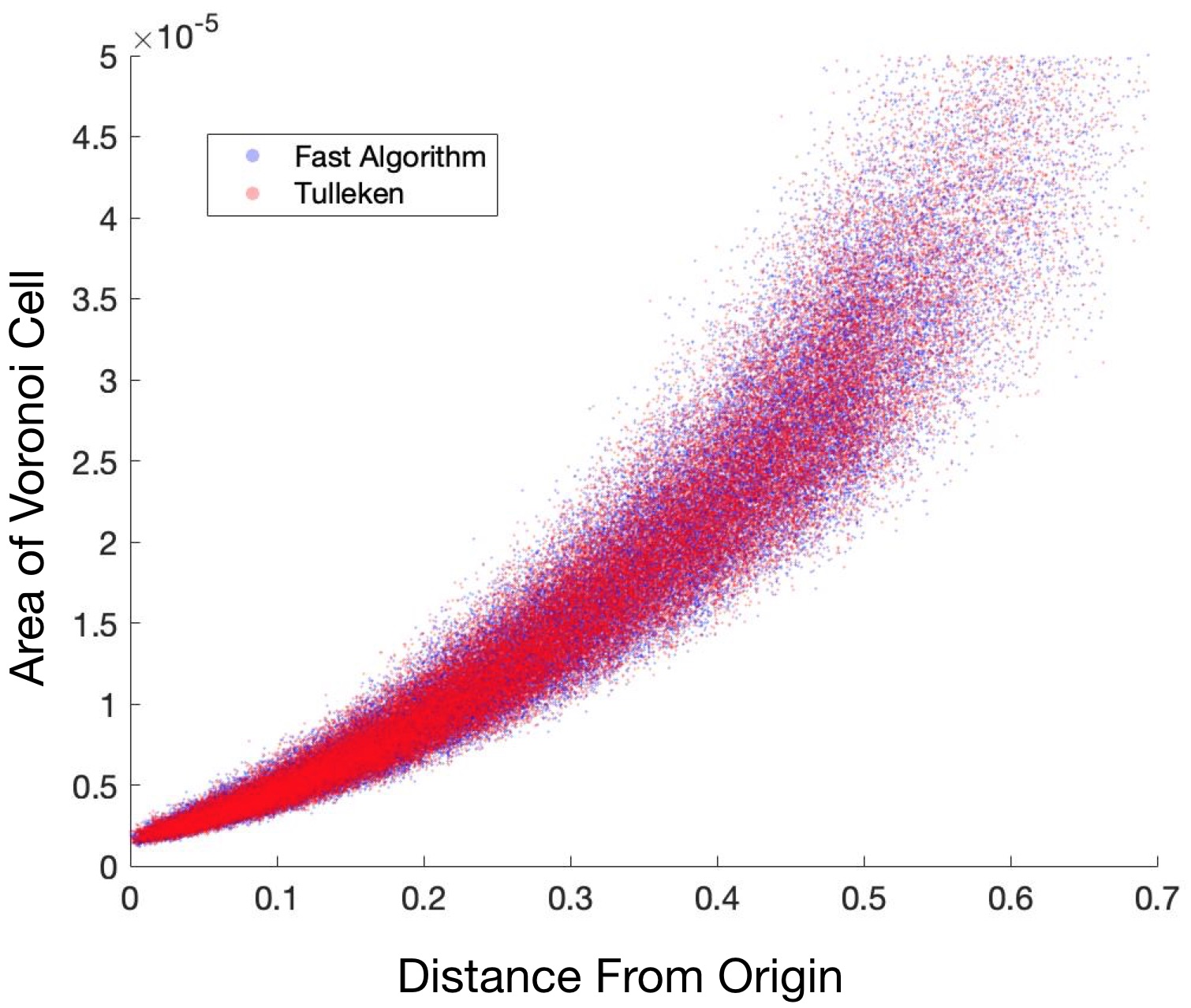}
  \end{center}
  \caption{ \protect\label{fig:cellAreas} Area of the Voronoi cell plotted against distance from the origin for each sample point with $\gamma=150$ for the fast algorithm (blue) and the Tulleken algorithm (red).  The distribution of points of the two algorithms is similar, indicating that they are generating sampling patterns of similar quality. }
\end{figure}

In Fig. \ref{fig:cellAreas}, we present a comparison of the results from Alg. \ref{alg:fastVDPD} to the variable density algorithm of Tulleken.  For the sampling patterns with $\gamma=150$ and without any additional directional undersampling, we created a Voronoi partition of the domain and plotted the area of each cell versus distance of the point from the origin.  Fig. \ref{fig:cellAreas} demonstrates that the distributions of area versus distance are very similar.

\begin{figure}[ht]
  \centering
  \includegraphics[width=0.98\linewidth]{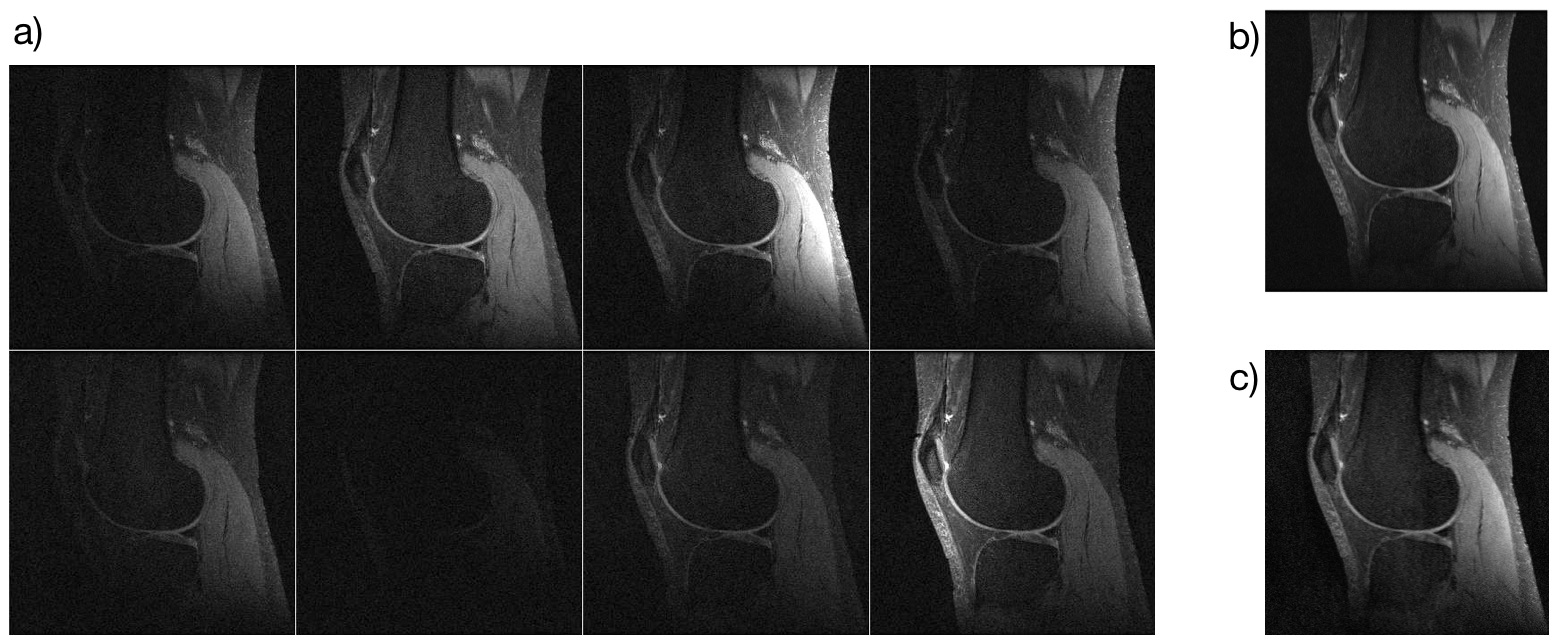}
  \caption{Fully sampled reconstructions of an 8-coil extremity coil.  The sensitivities of the coils in the anterior/posterior differ, whereas the sensitivity in the superior/inferior direction are approximately uniform for all coils. a) The fully sampled reconstructions for each of the 8 coils.  b) The sum-of-squares reconstruction with fully sampled data.  c) The SAKE+L1-ESPIRiT reconstruction of the data retrospectively subsampled with the sampling pattern of figure \ref{fig:vdpdSamplingPatterns} with $\gamma=125$ and undersampling of (3,1). }
  \label{fig:kneeRecons}
\end{figure}

\begin{figure}[ht]
  \centering
  \includegraphics[width=0.98\linewidth]{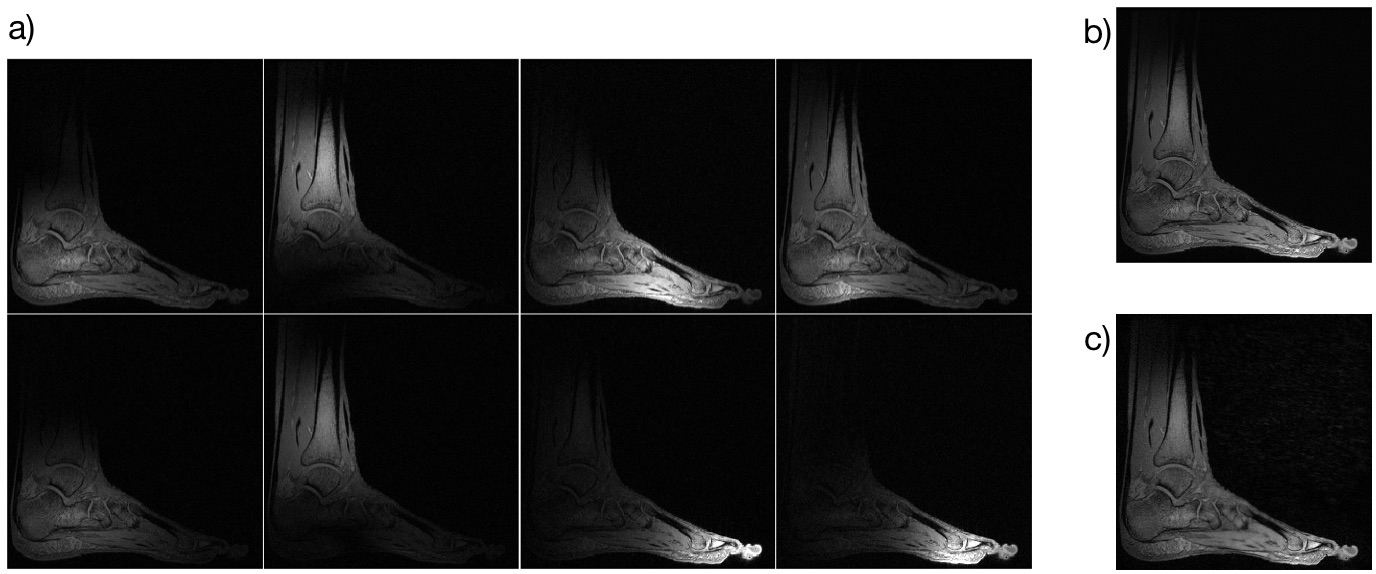}
  \caption{Fully sampled reconstructions of an 8-coil ankle coil.  a) The fully sampled reconstructions for each of the 8 coils.  b) The sum-of-squares reconstruction with fully sampled data.  c) The SAKE+L1-ESPIRiT reconstruction of the data retrospectively subsampled with the sampling pattern of figure \ref{fig:vdpdSamplingPatterns} with $\gamma=100$ and no directional undersampling. }
  \label{fig:ankleRecons}
\end{figure}

Figures \ref{fig:kneeRecons} and \ref{fig:ankleRecons} show how the sampling mask could be used with MRI data with knee and ankle data, respectively.  The data of Fig. \ref{fig:kneeRecons} was taken from \url{mridata.org} \cite{ong2018mridata}.  The data for Fig. \ref{fig:ankleRecons} was collected with a clinical 3 Tesla scanner and an 8-channel ankle coil.  Both of these datasets consist of fully sampled data with two dimensions of phase-encodes and a single dimension of readout.  An inverse Fast Fourier Transform was applied in the readout direction placing the data in a hybrid space; further processing was only done on a single slice.  The SAKE+L1-ESPIRiT algorithm was used to reconstruct data that was retrospectively subsampled with the variable density poisson disc sampling masks \cite{shin2014calibrationless,uecker2014espirit}.  Subfigure (a) shows the fully sampled reconstruction from each individual coil, (b) shows the sum-of-squares reconstruction with fully sampled data, and (c) shows the SAKE+L1-ESPIRiT reconstruction from retrospectively subsampled data.  The SAKE+L1-ESPIRiT reconstruction is very similar to the fully sampled sum-of-squares reconstruction.

\begin{figure}[ht]
  \begin{center}
    \includegraphics[width=0.95\linewidth]{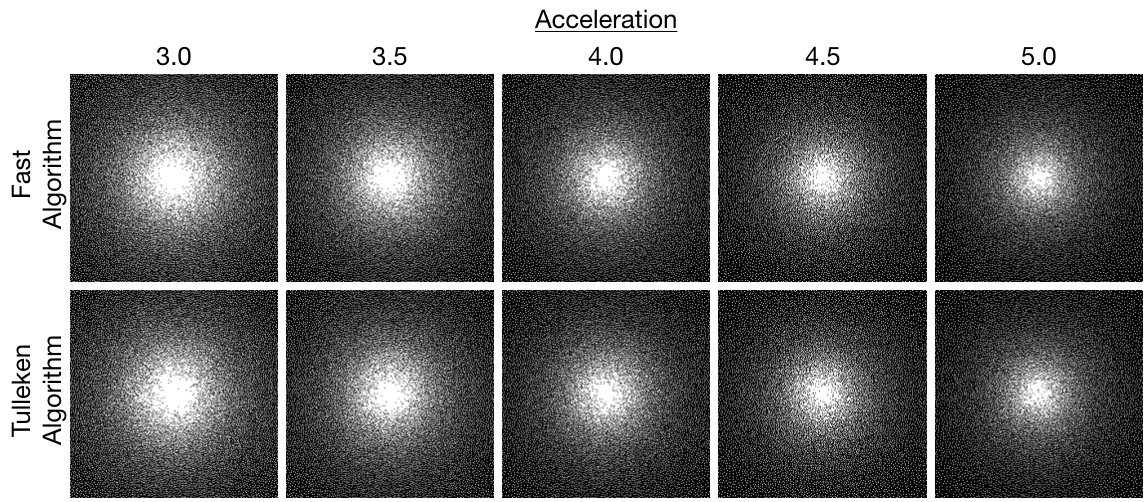}
  \end{center}
  \caption{ \protect\label{fig:accelerationPatterns} Acceleration Patterns }
\end{figure}

Figure \ref{fig:accelerationPatterns} shows patterns generated with specific accelerations using Alg. \ref{alg:binarySearch}.  Both the Tulleken and Alg. \ref{alg:fastVDPD} were used as the underlying method to determine the sampling pattern.  A tolerance of $0.01$ in the acceleration factor was permitted.  Indeed, sampling patterns with the specified acceleration rate were generated.  And, the sampling patterns of the Tulleken algorithm and Alg. \ref{alg:fastVDPD} are qualitatively similar.

In the spirit of reproducible research, we provide a software in C with implementations of the algorithms described. The software package can be downloaded from:  \url{https://github.com/ndwork/fastVDPD_C}.

\section{Discussion and Conclusion}
\label{sec:discussion}

In this work, we presented a fast algorithm for generating a variable density Poisson-disc sampling pattern, we show how the method can be adapted to permit a further directional undersampling, and we presented a method for generating a sampling pattern with a specific acceleration factor \cite{haldar2019oedipus}.  Having a fast algorithm for creating variable density Poisson-disc sampling patterns will empower future researchers to investigate different sampling distributions and determine the advantages and disadvantages of each one.  Additionally, different sampling patterns can be generated for different collections, which may be beneficial for some applications.

When selecting candidate points around an existing point, it may be the case that the candidates are unluckily chosen so that they conflict with existing points, but other candidate points could have been created that would not have.  Thus, the realized set of samples is not guaranteed to be maximal (where a maximal realization is one where the points are as dense as possible) \cite{ip2013pixelpie}.  However, the probability that the realization is maximal is typically very high (for sufficiently large values of $k$) \cite{johnsonDissertation2016}.

In this paper, we have shown that a simple scaling permits a computationally efficient method for generating patterns with directional undersampling.  A more general transformation can be used to create variations with more interesting patterns.  For example, an Affine transformation (or even any homeomorphism) could be used to generate sampling patterns with variable and directional sampling patterns.  The applications to MRI of such a technique are not immediately obvious.

It may be possible to further reduce the speed of sample generation by using parallelization to take advantage of multiple processing cores \cite{wei2008parallel}.  Additionally, it may be useful to adapt this algorithm to generate samples according to a variable density Poisson-disc distribution on a surface \cite{ying2013intrinsic,johnsonDissertation2016}.  We leave these prospects as possibilities for future work.

\section*{Acknowledgements}
ND and JP have been supported by the National Institute of Health's grant number P41 EB015891 and grant number R01 HL136965.
ND would like to thank the Quantitative Biosciences Institute at UCSF and the American Heart Association as funding sources for this work.



%% References



%% New version of the num-names style
%\bibliographystyle{unsrt}
%\bibliography{references.bib}

\begin{thebibliography}{10}

\bibitem{sodickson1997simultaneous}
Daniel~K Sodickson and Warren~J Manning.
\newblock Simultaneous acquisition of spatial harmonics ({SMASH}): fast imaging
  with radiofrequency coil arrays.
\newblock {\em Magnetic resonance in medicine}, 38(4):591--603, 1997.

\bibitem{pruessmann1999sense}
Klaas~P Pruessmann, Markus Weiger, Markus~B Scheidegger, and Peter Boesiger.
\newblock {SENSE}: sensitivity encoding for fast {MRI}.
\newblock {\em Magnetic resonance in medicine}, 42(5):952--962, 1999.

\bibitem{griswold2002generalized}
Mark~A Griswold, Peter~M Jakob, Robin~M Heidemann, Mathias Nittka, Vladimir
  Jellus, Jianmin Wang, Berthold Kiefer, and Axel Haase.
\newblock Generalized autocalibrating partially parallel acquisitions
  ({GRAPPA}).
\newblock {\em Magnetic Resonance in Medicine}, 47(6):1202--1210, 2002.

\bibitem{donoho2003optimally}
David~L Donoho and Michael Elad.
\newblock Optimally sparse representation in general (nonorthogonal)
  dictionaries via $\ell_1$ minimization.
\newblock {\em Proceedings of the National Academy of Sciences},
  100(5):2197--2202, 2003.

\bibitem{candes2006stable}
Emmanuel~J Candes, Justin~K Romberg, and Terence Tao.
\newblock Stable signal recovery from incomplete and inaccurate measurements.
\newblock {\em Communications on Pure and Applied Mathematics: A Journal Issued
  by the Courant Institute of Mathematical Sciences}, 59(8):1207--1223, 2006.

\bibitem{adcock2017breaking}
Ben Adcock, Anders~C Hansen, Clarice Poon, and Bogdan Roman.
\newblock Breaking the coherence barrier: A new theory for compressed sensing.
\newblock In {\em Forum of Mathematics, Sigma}, volume~5, pages 1--84.
  Cambridge University Press, 2017.

\bibitem{candes2008introduction}
Emmanuel~J Cand{\`e}s and Michael~B Wakin.
\newblock An introduction to compressive sampling [a sensing/sampling paradigm
  that goes against the common knowledge in data acquisition].
\newblock {\em {IEEE} signal processing magazine}, 25(2):21--30, 2008.

\bibitem{lustig2007sparse}
Michael Lustig, David Donoho, and John~M Pauly.
\newblock Sparse {MRI}: The application of compressed sensing for rapid {MR}
  imaging.
\newblock {\em Magnetic Resonance in Medicine}, 58(6):1182--1195, 2007.

\bibitem{vasanawala2011practical}
SS~Vasanawala, MJ~Murphy, Marcus~T Alley, P~Lai, Kurt Keutzer, John~M Pauly,
  and Michael Lustig.
\newblock Practical parallel imaging compressed sensing {MRI}: Summary of two
  years of experience in accelerating body {MRI} of pediatric patients.
\newblock In {\em 2011 ieee international symposium on biomedical imaging: From
  nano to macro}, pages 1039--1043. IEEE, 2011.

\bibitem{gamper2008compressed}
Urs Gamper, Peter Boesiger, and Sebastian Kozerke.
\newblock Compressed sensing in dynamic {MRI}.
\newblock {\em Magnetic Resonance in Medicine}, 59(2):365--373, 2008.

\bibitem{cline2009dart}
David Cline, Stefan Jeschke, K~White, Anshuman Razdan, and Peter Wonka.
\newblock Dart throwing on surfaces.
\newblock In {\em Computer Graphics Forum}, volume 28 (4), pages 1217--1226.
  Wiley Online Library, 2009.

\bibitem{lagae2008comparison}
Ares Lagae and Philip Dutr{\'e}.
\newblock A comparison of methods for generating poisson disk distributions.
\newblock In {\em Computer Graphics Forum}, volume 27 (1), pages 114--129.
  Wiley Online Library, 2008.

\bibitem{bridson2007fast}
Robert Bridson.
\newblock Fast poisson disk sampling in arbitrary dimensions.
\newblock In {\em {SIGGRAPH} sketches}, page~22, 2007.

\bibitem{tulleken2008poisson}
Herman Tulleken.
\newblock Poisson disk sampling.
\newblock {\em Dev. Mag}, 21:21--25, 2008.

\bibitem{levine20173d}
Evan Levine, Bruce Daniel, Shreyas Vasanawala, Brian Hargreaves, and Manojkumar
  Saranathan.
\newblock {3D} cartesian {MRI} with compressed sensing and variable view
  sharing using complementary poisson-disc sampling.
\newblock {\em Magnetic resonance in medicine}, 77(5):1774--1785, 2017.

\bibitem{noll1991homodyne}
Douglas~C Noll, Dwight~G Nishimura, and Albert Macovski.
\newblock Homodyne detection in magnetic resonance imaging.
\newblock {\em {IEEE} transactions on medical imaging}, 10(2):154--163, 1991.

\bibitem{ong2018mridata}
F~Ong, S~Amin, S~Vasanawala, and M~Lustig.
\newblock Mridata.org: An open archive for sharing {MRI} raw data.
\newblock In {\em Proc. Intl. Soc. Mag. Reson. Med}, volume~26, page~1, 2018.
\newblock \url{www.mridata.org}.

\bibitem{shin2014calibrationless}
Peter~J Shin, Peder~EZ Larson, Michael~A Ohliger, Michael Elad, John~M Pauly,
  Daniel~B Vigneron, and Michael Lustig.
\newblock Calibrationless parallel imaging reconstruction based on structured
  low-rank matrix completion.
\newblock {\em Magnetic resonance in medicine}, 72(4):959--970, 2014.

\bibitem{uecker2014espirit}
Martin Uecker, Peng Lai, Mark~J Murphy, Patrick Virtue, Michael Elad, John~M
  Pauly, Shreyas~S Vasanawala, and Michael Lustig.
\newblock {ESPIRiT}—an eigenvalue approach to autocalibrating parallel {MRI}:
  where {SENSE} meets {GRAPPA}.
\newblock {\em Magnetic resonance in medicine}, 71(3):990--1001, 2014.

\bibitem{haldar2019oedipus}
Justin~P Haldar and Daeun Kim.
\newblock {OEDIPUS}: An experiment design framework for sparsity-constrained
  {MRI}.
\newblock {\em IEEE transactions on medical imaging}, 38(7):1545--1558, 2019.

\bibitem{ip2013pixelpie}
Cheuk~Yiu Ip, M~Adil Yal{\c{c}}in, David Luebke, and Amitabh Varshney.
\newblock Pixelpie: maximal poisson-disk sampling with rasterization.
\newblock In {\em Proceedings of the 5th High-Performance Graphics Conference},
  pages 17--26. ACM, 2013.

\bibitem{johnsonDissertation2016}
Ethan M.~I. Johnson.
\newblock {\em Techniques and Analyses for Ultrashort Echo-time Magnetic
  Resonance Imaging}.
\newblock PhD thesis, Stanford University, 2016.

\bibitem{wei2008parallel}
Li-Yi Wei.
\newblock Parallel poisson disk sampling.
\newblock {\em ACM Transactions on Graphics (TOG)}, 27(3):20, 2008.

\bibitem{ying2013intrinsic}
Xiang Ying, Shi-Qing Xin, Qian Sun, and Ying He.
\newblock An intrinsic algorithm for parallel poisson disk sampling on
  arbitrary surfaces.
\newblock {\em {IEEE} transactions on visualization and computer graphics},
  19(9):1425--1437, 2013.

\end{thebibliography}

\end{document}